\documentclass[aps,pra,superscriptaddress,floatfix,twocolumn]{revtex4}
\usepackage[dvips]{graphicx}

\usepackage{longtable}
\usepackage{dcolumn}
\usepackage[dvips]{graphicx}
\usepackage{bm}
\usepackage{lipsum}
\usepackage{bbm}
\usepackage{color}

\usepackage{times}
\usepackage{nicefrac}
\usepackage{amsmath}
\usepackage{amsfonts}
\usepackage{amssymb}
\usepackage{amsthm}
\usepackage[normalem]{ulem}
\newcolumntype{.}{D{x}{}{-1}}
\usepackage{tikz}
\usetikzlibrary{snakes}
\usepackage{orcidlink}

\def\ketm#1{  \left\vert  #1   \right\rangle   }

\def\sprm#1#2{  \left\langle #1 \left\vert \right. #2 \right\rangle   }

\def\mem#1#2#3{  \left\langle #1 \left\vert  #2 \right\vert #3 \right\rangle   }

\DeclareMathOperator*{\SumInt}{%
\mathchoice%
  {\ooalign{$\displaystyle\sum$\cr\hidewidth$\displaystyle\int$\hidewidth\cr}}
  {\ooalign{\raisebox{.14\height}{\scalebox{.7}{$\textstyle\sum$}}\cr\hidewidth$\textstyle\int$\hidewidth\cr}}
  {\ooalign{\raisebox{.2\height}{\scalebox{.6}{$\scriptstyle\sum$}}\cr$\scriptstyle\int$\cr}}
  {\ooalign{\raisebox{.2\height}{\scalebox{.6}{$\scriptstyle\sum$}}\cr$\scriptstyle\int$\cr}}
}

\begin{document}

\title{Bound-state Compton scattering of linearly polarized photons}

\author{Jonas~Sommerfeldt, \orcidlink{0000-0002-3471-7494}}
\affiliation{Laboratoire Kastler Brossel, Sorbonne Universit\'e, CNRS, ENS-Universit\'e PSL, Coll\'ege de France, Campus Pierre et Marie Curie, F-75005 Paris, France}
\affiliation{Max Planck Institute for Nuclear Physics, D-69117 Heidelberg, Germany}

\author{Nick~M.~Mayer, \orcidlink{0009-0005-9188-6277}}
\affiliation{Physikalisch-Technische Bundesanstalt, D-38116 Braunschweig, Germany}
\affiliation{Institut f\"ur Mathematische Physik, Technische Universit\"at Braunschweig, D-38106 Braunschweig, Germany}

\author{Anna Maiorova, \orcidlink{0000-0002-7385-5582}}
\affiliation{GSI Helmholtzzentrum für Schwerionenforschung, D-64291 Darmstadt, Germany}
\affiliation{Helmholtz Institute Jena, D-07743 Jena, Germany}

\author{Wilko Middents, \orcidlink{0000-0001-8105-4657}}
\affiliation{GSI Helmholtzzentrum für Schwerionenforschung, D-64291 Darmstadt, Germany}
\affiliation{Helmholtz Institute Jena, D-07743 Jena, Germany}
\affiliation{Institut f\"ur Optik und Quantenelektronik, Friedrich-Schiller-Universit\"at Jena, D-07743 Jena, Germany}

\author{Stephan Fritzsche, \orcidlink{0000-0003-3101-2824}}
\affiliation{GSI Helmholtzzentrum für Schwerionenforschung, D-64291 Darmstadt, Germany}
\affiliation{Helmholtz Institute Jena, D-07743 Jena, Germany}
\affiliation{Theoretisch-Physikalisches Institut, Friedrich-Schiller-Universit\"at Jena, D-07743 Jena, Germany}

\author{Thomas St\"ohlker, \orcidlink{0000-0003-0461-3560}}
\affiliation{GSI Helmholtzzentrum für Schwerionenforschung, D-64291 Darmstadt, Germany}
\affiliation{Helmholtz Institute Jena, D-07743 Jena, Germany}
\affiliation{Institut f\"ur Optik und Quantenelektronik, Friedrich-Schiller-Universit\"at Jena, D-07743 Jena, Germany}

\author{Andrey~Surzhykov. \orcidlink{0000-0002-6441-0864}}
\affiliation{Physikalisch-Technische Bundesanstalt, D-38116 Braunschweig, Germany}
\affiliation{Institut f\"ur Mathematische Physik, Technische Universit\"at Braunschweig, D-38106 Braunschweig, Germany}

\begin{abstract}
We present a theoretical study of Compton scattering of X- and $\gamma$-rays by a $K$-shell electron. Special attention is paid to the double-differential cross section and polarization of the scattered photons for linearly polarized incident photons. To investigate these observables, we employ the scattering matrix (S-matrix) approach based on relativistic Green's functions. The S-matrix results are moreover compared with predictions of the free-electron and impulse approximations, allowing us to assess the role of electron binding effects. Detailed calculations are carried out for hydrogen-like Ne$^{9+}$ and Pb$^{81+}$ targets over a wide range of incident photon energies and scattering angles. The calculations reveal kinematic regimes in which the impulse approximation agrees reasonably well with the S-matrix results. We also explore the polarization of scattered photons for slightly depolarized incident radiation, including the highly sensitive case of scattering at $90^\circ$.
\end{abstract}
\maketitle
\section{Introduction}

Compton scattering of a photon by an electron, initially bound to an atom or an ion, is one of the fundamental processes of light-matter interaction. In this process, the photon is generally scattered with lower frequency, while part of its energy is used to eject the electron, making this process one of the key mechanisms of energy transfer from electromagnetic radiation to matter. Since the seminal papers by Arthur Compton \cite{Com23a, Com23b}, which earned him the Nobel Prize in Physics in 1927, this photon scattering has received much attention in many areas of modern research.  In particular, Compton scattering finds important applications in medicine, playing a crucial role in radiobiology, radiation therapy, medical imaging, and tomographic systems \cite{NorJAP94}. It also allows one to probe the electron momentum density in solids \cite{Coo97,AGUIAR201564}, making this process a powerful tool for material science research. Based on the Compton effect, novel segmented semiconductor detectors have been developed to measure the polarization of hard X-rays \cite{Spi08,Voc17}, which are important for gamma spectroscopy. Moreover, the Compton effect is widely employed in astrophysics as well as in atomic and plasma physics.

Earlier investigations of Compton scattering have focused mainly on the cross sections and energy distributions of the scattered radiation. In recent years, however, there has been a growing interest in the \textit{linear polarization} of scattered photons, especially when the incident radiation is itself polarized. Such studies have been stimulated by the availability of highly polarized X-ray beams at modern synchrotron facilities, such as PETRA III at DESY in Hamburg, as well as by the development of polarization-sensitive detectors \cite{WeB15, BeC22}. The analysis of experimental data on the polarization of Compton-scattered radiation requires the development of reliable theoretical models. Since the discovery of the Compton effect, several theoretical approaches capable of describing polarization effects have been developed, each with its own range of validity and accuracy. The simplest approach is provided by the free-electron approximation (FEA), which is based on the Klein–Nishina formula for photon scattering by a free electron at rest \cite{KleNish1929}. In this case, the energy of the scattered photon is uniquely determined by the scattering angle, giving rise to a well-defined Compton peak in the energy spectrum. Due to its simplicity, the FEA is widely used as a first approximation in Compton scattering, particularly at high photon energies \cite{PrL10}. However, at lower energies or for heavier targets, atomic binding effects become significant, and more refined models such as the impulse approximation (IA) must be employed \cite{DuM29,Rib75,RibII75,BERGSTROM19973,QiC20,QiW21,MiG25}. Within the IA framework, a bound electron is approximated as quasi-free one; however, it no longer remains at rest in the laboratory frame, and its motion is described by a certain momentum distribution. For sufficiently large photon momentum transfer, the impulse approximation reproduces the differential cross section well, especially near the Compton peak \cite{PRATT2010124,QiC20,QiW21}. Unlike cross sections, the polarization of the scattered photons is generally more sensitive to the details of the interaction, especially to electron binding effects. Its quantitative analysis therefore often requires a more rigorous theoretical framework such as the relativistic S-matrix approach.

In this paper, we present a relativistic S-matrix treatment of Compton scattering of linearly polarized photons by bound electrons. Particular emphasis is placed on the double-differential cross section if the emitted electron remains unobserved and, especially, on the linear polarization of the scattered photons. Before analyzing the Compton cross section and polarization properties, we specify the geometry of the scattering process in Sec.~\ref{geometry}. A brief overview of the theoretical approaches used to describe this process, with special emphasis on the S-matrix theory, is presented in Sec.~\ref{theory}. We stress, in particular, that the S-matrix treatment requires a summation over the complete spectrum of the atomic Hamiltonian, including both bound and continuum states. This summation can be carried out most efficiently using the relativistic Green's function approach discussed in Sec.~\ref{sec:theory_green}. The results of the S-matrix calculations for the $K$-shell Compton scattering by Ne$^{9+}$ and Pb$^{81+}$ ions are presented in Sec.~\ref{sec:results} and compared with the predictions of the free-electron and impulse approximations. Calculations have been carried out for a broad range of kinematic regimes characterized by the ratio of the bound-electron momentum $p_b$ to the photon momentum transfer $q$. For $p_b/q \lesssim 1$ and photon energies near the Compton peak, the S-matrix predictions generally agree well with those of the impulse and free-electron approximations. While such agreement for the cross sections has been well known \cite{BeS93}, we find that it also holds for the polarization of the scattered photons. In addition, we investigate how the polarization of the scattered photons changes for slightly depolarized incident radiation. In analogy to Rayleigh scattering \cite{SuY13,BlF16}, a particularly strong sensitivity to initial polarization is observed for scattering at angles near $90^\circ$. Finally, Sec.~\ref{sec:summary} provides a summary of the results and a brief outlook.

Relativistic units ( $\hbar = c = m_e = 1$) are used throughout the paper.

\section{Geometry and basic notations} 
\label{geometry}

\begin{figure}[t] 
    \includegraphics[width=\linewidth]{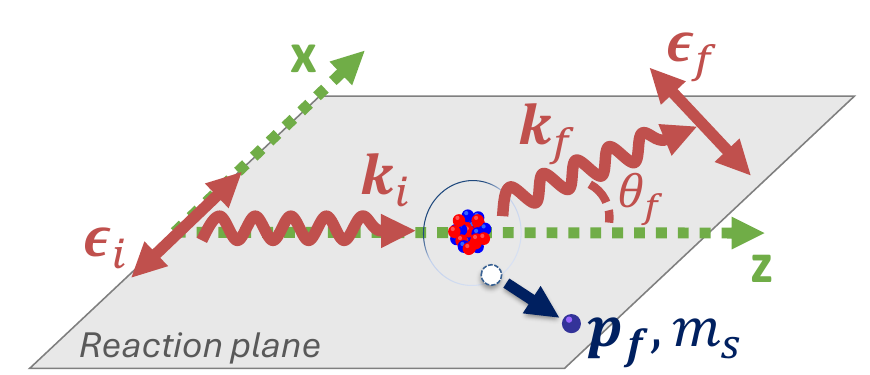}
    \caption{Geometry of the bound-state Compton scattering. The quantization ($z$-) axis is chosen along the wave vector ${\bm k}_i$ of the incident photon, which, together with the wave vector ${\bm k}_f$ of the scattered photon, defines the reaction ($xz$-) plane.}
    \label{Fig1_geometry}
\end{figure}

Before discussing the theoretical approaches to the description of bound-state Compton scattering, let us briefly recall the geometry of the process and the basic notations.  In this paper, we consider the scattering of an incident photon with energy $\omega_i$, wave vector ${\bm k}_i$ and polarization ${\bm \epsilon}_i$ by the $K$-shell electron. The energy as well as the wave- and polarization vectors of the scattered photon are $\omega_f$, ${\bm k}_f$ and ${\bm \epsilon}_f$, respectively, while ${\bm p}_f$ and $m_s$ denote the asymptotic momentum and spin projection of the emitted electron, which remain unobservable in our study. It is implied that the wave vector ${\bm k}_i$ of the incident photon defines the quantization ($z$-) axis and, together with ${\bm k}_f$ establishes the reaction ($xz$-) plane, see Fig.~\ref{Fig1_geometry}. For this geometry, the emission of the scattered photon is described by a single polar angle $\theta_f$. The linear polarization vectors ${\bm \epsilon}_i$ and ${\bm \epsilon}_f$ lie in planes perpendicular to the corresponding photon momenta ${\bm k}_i$ and ${\bm k}_f$, respectively.  

\section{Theoretical background} 
\label{theory}

\subsection{Cross sections and polarization parameters}
\label{subsec:theory_polarization}

Since the present study is primarily concerned with the \textit{linear polarization} of incident and scattered photons, let us first recall the way to describe it. As is usual in optics and atomic physics, for this purpose we will use the Stokes parameters, which completely specify both the degree and direction of polarization of a photon. 
The first Stokes parameter
\begin{equation}
   \label{eq:Stokes_P1}
   P_1 = \frac{I_{0} - I_{90}}{I_{0} + I_{90}}
\end{equation}
is related to the intensities $I_{0}$ and $I_{90}$ of radiation linearly polarized parallel and perpendicular to the reaction ($xz$-) plane, respectively. The second Stokes parameter $P_2$ is defined analogously, but in terms of the intensities $I_{45}$ and $I_{135}$, corresponding to linear polarization at angles of 45$^\circ$ and 135$^\circ$ with respect to the $xz$-plane. The third Stokes parameter $P_3$, which describes circular polarization, is assumed to be zero in the present study and will therefore not be considered further.

In what follows, we will assume that the polarization of incident photons is described by the parameters $0 \ll P_1^{(i)} \leq 1$ and $P_2^{(i)} = 0$. This corresponds to a rather common scenario of synchrotron experiments, where the incoming photons are strongly (though not always completely) linearly polarized within the reaction plane \cite{BlF16}.

To determine the linear polarization of the scattered photons, one must start from the Compton cross section. As mentioned above, we assume that the emitted electron is not observed. Consequently, the cross section is doubly differential in the energy $\omega_f$ and angle $\theta_f$ of the scattered photon. For the case of linearly polarized incident radiation, this cross section can be written as:
\begin{eqnarray}
   \label{eq:cross_sections_general}
   \frac{{\rm d^2}\sigma}{{\rm d}\Omega_f {\rm d}\omega_f}\left({\bm \epsilon_f},  P_1^{(i)}\right) &=& \frac{1 + P_1^{(i)}}{2} \frac{{\rm d^2}\sigma}{{\rm d}\Omega_f {\rm d}\omega_f}\left({\bm \epsilon_f}, {\bm \epsilon}_{i,\parallel} \right) \nonumber \\
   &+& \frac{1 - P_1^{(i)}}{2} \frac{{\rm d^2}\sigma}{{\rm d}\Omega_f {\rm d}\omega_f}\left({\bm \epsilon_f}, {\bm \epsilon}_{i, \perp} \right)  ,
\end{eqnarray}
since we took $P_2^{(i)} = 0$. In this expression, the cross sections
\begin{eqnarray}
    \frac{{\rm d^2}\sigma}{{\rm d}\Omega_f {\rm d}\omega_f}\left({\bm \epsilon}_{f}, {\bm \epsilon}_{i,\parallel} \right) \equiv \frac{{\rm d^2}\sigma}{{\rm d}\Omega_f {\rm d}\omega_f}\left({\bm \epsilon}_{f}, P_1^{(i)}=1 \right) \, , \nonumber \\ 
    \frac{{\rm d^2}\sigma}{{\rm d}\Omega_f {\rm d}\omega_f}\left({\bm \epsilon}_{f}, {\bm \epsilon}_{i,\perp} \right) \equiv \frac{{\rm d^2}\sigma}{{\rm d}\Omega_f {\rm d}\omega_f}\left({\bm \epsilon}_{f}, P_1^{(i)}=-1 \right)
\end{eqnarray}
describe the scattering of the photon with the polarization vector ${\bm \epsilon_f}$ for the case where the incident photon has the polarization ${\bm \epsilon}_{i}$ along the $x$- and $y$- axes, respectively.

Using Eqs.~(\ref{eq:Stokes_P1}) and (\ref{eq:cross_sections_general}) as well as the natural assumption that the intensity of the scattered light is proportional to the corresponding Compton cross section, we can obtain the first Stokes parameter for the final-state photons as:
\begin{eqnarray}
    \label{eq:Stokes_P1_final}
    P^{(f)}_1 &=& \frac{{\rm d^2}\sigma\left({\bm \epsilon}_{f, \parallel},  P_1^{(i)}\right) - {\rm d^2}\sigma\left({\bm \epsilon}_{f, \perp},  P_1^{(i)}\right)}{{\rm d^2}\sigma\left({\bm \epsilon}_{f,\parallel},  P_1^{(i)}\right) + {\rm d^2}\sigma\left({\bm \epsilon}_{f,\perp},  P_1^{(i)}\right)} \, ,
\end{eqnarray}
while the second one is $P^{(f)}_2 = 0$, and, for brevity, we have introduced the shorthand notation ${\rm d^2}\sigma \equiv {\rm d^2}\sigma/{\rm d}\Omega_f {\rm d}\omega_f$ that will be used later in the text. Apparently, any further analysis of the polarization $P^{(f)}_1$ requires knowledge about the \textit{polarization-resolved} Compton cross sections ${\rm d^2}\sigma\left({\bm \epsilon_f}, {\bm \epsilon_i} \right)$. In the following sections, therefore, we will discuss their evaluation within the free-electron approach, impulse approximation, and relativistic S-matrix theory.

\subsection{Free electron approximation}
\label{subsec:theory_FEA}

Not much needs to be said about the Compton scattering of a photon by a free electron at rest. The double-differential cross section for this case is given by the well-known Klein-Nishina formula:
\begin{eqnarray}
    \label{eq:Klein_Nishina_formula}
    \frac{{\rm d^2}\sigma^{\rm (FEA)}}{{\rm d}\Omega_f {\rm d}\omega_f}\left({\bm \epsilon_f}, {\bm \epsilon_i} \right) &=& \frac{r_0^2}{4} \, \left( \frac{\omega_f}{\omega_i} \right)^2 \nonumber \\[0.2cm]
    & & \hspace*{-2.3cm} \times \left[ \frac{\omega_f}{\omega_i} + \frac{\omega_i}{\omega_f} + 4 \left|{\bm \epsilon}^*_i \cdot {\bm \epsilon}_f \right|^2 - 2\right] \, \delta\left(\omega_f - \omega_C \right) \, ,
\end{eqnarray}
where $r_0$ is the classical electron radius and the Compton energy is
\begin{equation}
    \label{eq:Compton_energy}
    \omega_C = \frac{\omega_i}{1 + \omega_i \left(1 - \cos\theta_f \right)} \, .
\end{equation}
The last expression implies that, for a given angle $\theta_f$, the scattered photon energy is uniquely determined.

As seen from Eq.~(\ref{eq:Klein_Nishina_formula}), the polarization dependence of the Klein-Nishina cross section is introduced by means of the scalar product ${\bm \epsilon}^*_i \cdot {\bm \epsilon}_f$. For the geometry shown in Fig.~\ref{Fig1_geometry}, the square of this product reads:
\begin{eqnarray}
    \label{eq:polarization_vectors_product}
    \left| {\bm \epsilon}^*_{i, \parallel} \cdot {\bm \epsilon}_{f, \parallel} \right|^2 &=& \cos^2\theta_f \, , \nonumber \\[0.2cm]
    \left| {\bm \epsilon}^*_{i, \perp} \cdot {\bm \epsilon}_{f, \perp} \right|^2  &=& 1, \nonumber \\[0.2cm]
    \left|{\bm \epsilon}^*_{i, \parallel} \cdot {\bm \epsilon}_{f, \perp}\right|^2 &=& \left|{\bm \epsilon}^*_{i, \perp} \cdot {\bm \epsilon}_{f, \parallel}\right|^2 = 0 \, , \\ \nonumber
\end{eqnarray}
for different combinations of polarization vector directions. By inserting the expressions (\ref{eq:polarization_vectors_product}) into Eq.~(\ref{eq:Stokes_P1_final}) we derive the polarization of scattered photon for the free-electron case:
\begin{eqnarray}
    \label{eq:Stokes_P1_final_FEA}
    P^{(f)}_1 &=& \frac{2P^{(i)}_1 - \left(1 + P^{(i)}_1\right) \sin^2\theta_f}{\left(\omega_i^2 + \omega_f^2\right)/\omega_f \omega_i - \left(1 + P^{(i)}_1\right) \sin^2\theta_f} \, ,
\end{eqnarray}
where $\omega_f = \omega_C$. 

For the analysis below, it is useful to recall the classical Thomson limit of Compton scattering by a free electron at rest. Indeed, when the incident photon energy is much smaller than the electron rest energy, $\omega_i \ll 1$, the electron recoil can be neglected. Equation (\ref{eq:Compton_energy}) then yields $\omega_C \approx \omega_i$ so that the scattered photon energy is approximately equal to the incident one. In this limit, the cross section (\ref{eq:Klein_Nishina_formula}) reduces to the classical expression:
\begin{eqnarray}
    \label{eq:Thomson}
    \frac{{\rm d^2}\sigma^{\rm (Th)}}{{\rm d}\Omega_f {\rm d}\omega_f}\left({\bm \epsilon_f}, {\bm \epsilon_i} \right) &\approx& r_0^2 \, \left|{\bm \epsilon}^*_i \cdot {\bm \epsilon}_f \right|^2 \, \delta\left(\omega_f - \omega_i \right) \, .
\end{eqnarray}
The above expression shows that the Thomson cross section depends explicitly on the mutual orientation of the polarization vectors of the incident and scattered photons. This feature will be useful below for a qualitative analysis of the polarization dependence of the scattering process.

\subsection{Impulse approximation}
\label{subsec:theory_impulse_approximation}

Since the impulse approximation (IA) has already been widely discussed in the literature \cite{Rib75,QiC20,QiW21,MiG25}, we restrict ourselves here to a brief review of its main ideas. While the basic formulas (\ref{eq:Klein_Nishina_formula})--(\ref{eq:Compton_energy}) of the free-electron model are derived for an electron at rest, the impulse approximation accounts for the momentum distribution of an electron initially bound in a target atom. In this approach, the bound electron is treated as quasi-free, but with a nonzero initial momentum described by the bound-state momentum distribution. The IA double-differential scattering cross section can then be written as:
\begin{eqnarray}
   \label{eq:cross_section_IA}
   \frac{{\rm d^2}\sigma^{\rm (IA)}}{{\rm d}\Omega_f {\rm d}\omega_f}\left({\bm \epsilon_f}, {\bm \epsilon_i} \right) && \nonumber \\[0.2cm]
   && \hspace{-2.2cm}= \int \frac{{\rm d^2}\sigma^{\rm (FEA)}}{{\rm d}\Omega_f {\rm d}\omega_f}\left({\bm \epsilon_f}, {\bm \epsilon_i}; {\bm p}_i \right) \,
   \rho\left( {\bm p}_i \right) \, \delta\left(\Delta E \right) \, {\rm d}{\bm p}_i ,
\end{eqnarray}
upon convoluting its free-electron counterpart ${\rm d^2}\sigma^{\rm (FEA)}\left({\bm \epsilon_f}, {\bm \epsilon_i}; {\bm p}_i \right)$ with the momentum distribution $\rho({\bm p}_i)$ of a ``bound'' electron. In Eq.~(\ref{eq:cross_section_IA}), moreover, the Dirac delta function ensures energy conservation:
\begin{equation}
    \label{eq:energy_conservation}
    \Delta E = E_i + \omega_i - E_f - \omega_f = 0 \, ,
\end{equation}
where $E_i$ and $E_f$ are the energies of the initial (bound) and final (continuum) electron.

\begin{figure}[t]
    \includegraphics[width=5.8cm]{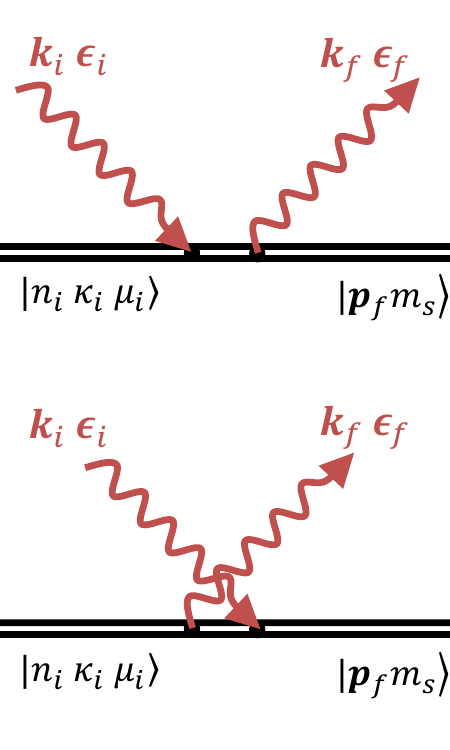}
    \caption{The Feynman diagrams that describe the Compton scattering. The wavy lines represent incident and outgoing photons while the double straight line --- the electron, moving in the field of a nucleus. The upper and lower diagrams are usually referred to as ``absorption first'' and ``emission first'', respectively.}
     \label{Fig2_diagmrams}
\end{figure}

The FEA cross section in Eq.~(\ref{eq:cross_section_IA}) is obtained by performing a Lorentz transformation of the Klein–Nishina formula (\ref{eq:Klein_Nishina_formula}), which is defined in the electron rest frame, to the laboratory frame. In the latter, the quasi-free electron carries an initial momentum ${\bm p}_i = (|p_i|\sin\psi\cos\varphi,|p_i|\sin\psi\sin\varphi,|p_i|\cos\psi)$ which, in general, does not lie in the reaction ($xz$-) plane. In this case, the doubly differential cross section can be written as:
%
%
\begin{eqnarray}
    \label{eq:Klein_Nishina_formula_mod}
    \frac{{\rm d^2}\sigma^{\rm (FEA)}}{{\rm d}\Omega_f {\rm d}\omega_f}\left({\bm \epsilon_f}, {\bm \epsilon_i}; {\bm p}_i \right) &=& \frac{\partial (\omega'_f, \Omega'_f)}{\partial (\omega_f, \Omega_f)}\,\frac{r_0^2}{4} \, \left( \frac{\omega'_f}{\omega'_i} \right)^2 \nonumber \\[0.2cm]
    & & \hspace*{-3.2cm} \times \left[ \frac{\omega'_f}{\omega'_i} + \frac{\omega'_i}{\omega'_f} + 4 \left|{\bm \epsilon}^*_i \cdot {\bm \epsilon}_f \right|^2 - 2\right] \, \delta\left(\omega'_f - \omega'_C \right),
\end{eqnarray}
where the Jacobian $\partial (\omega'_f, \Omega'_f)/\partial (\omega_f, \Omega_f)$ accounts for the transformation from the electron rest frame (denoted here by primes) to the laboratory frame, see, e.g., \cite{EICHLER20071}. The energies of the incident and scattered photons in the electron rest frame are given by $\omega'_i =\omega_i \gamma\left(1-\beta\cos\psi\right)$ and $\omega'_f =\omega_f\gamma\left(1-\beta\left(\sin\theta_f\sin\psi\cos\varphi+\cos\theta_f\cos\psi\right)\right)=\omega_f\gamma\left(1-\beta\cos\zeta\right)$, respectively. Here, $\beta = v_i$ and $\gamma = 1/\sqrt{1-\beta^2}$ are the standard relativistic parameters. The angle $\zeta$ denotes the angle between the scattered photon and the incident electron in the laboratory frame. Finally, the Compton energy in the laboratory frame reads:
\begin{eqnarray}
    \label{eq:omega_f_me}
    \omega^{\rm (lab)}_C &=& \frac{\omega_i \gamma\left(1-\beta\cos{\psi}\right)}{\omega_i\left(1-\cos{\theta_f}\right)+\gamma\left(1-\beta\cos{\zeta}\right)} \, .
\end{eqnarray}
%
%


Using Eqs.~(\ref{eq:cross_section_IA}) and (\ref{eq:Klein_Nishina_formula_mod}) together with an appropriate model for the electron momentum distribution $\rho({\bm p}_i)$, one can calculate the double-differential cross sections ${\rm d^2}\sigma^{\rm (IA)}\!\left({\bm \epsilon_f}, {\bm \epsilon_i}\right)$ and, consequently, the linear polarization of the scattered photons. In the present work, the momentum distribution $\rho({\bm p}_i)$ is obtained from the Fourier transform of the bound-state Dirac wave functions for hydrogen-like Ne$^{9+}$ and Pb$^{81+}$ ions in their ground ($1s$) state. The integration over the initial electron momentum in Eq.~(\ref{eq:cross_section_IA}) was performed using the Monte Carlo procedure described in Ref.~\cite{MiG25}.

Having discussed the basic formulas of the impulse approximation, we now briefly discuss its range of validity. A convenient criterion for the validity of the IA is provided by the ratio of the characteristic momentum $p_b = \alpha Z$ of a bound electron to the photon momentum transfer $q = |\bm{k}_i - \bm{k}_f|$. In the regime $p_b/q < 1$, the IA is known to provide a good approximation to the (unpolarized) double-differential cross section, particularly for energies of scattered photons near the Compton peak \cite{BeS93,PRATT2010124}. For $p_b/q > 1$, or when the energy transfer becomes comparable to the binding energy, electron binding effects play an increasingly important role, and the accuracy of IA  deteriorates.

\subsection{Scattering matrix theory}
\label{subsec:S_matrix}

A more rigorous, compared to FEA and IA, approach to the description of Compton scattering is based on the use of the relativistic scattering matrix. Within the S-matrix approach, the electron in its initial and final states is described by the bound $\ketm{n_i \kappa_i \mu_i}$ and continuum $\ketm{{\bm p}_f m_s}$ solutions of the Dirac equation for a Coulomb (or, if necessary, mean-field screening) potential. While the continuum electron is characterized by the asymptotic momentum ${\bm p}_f$ and spin projection $m_s$ onto its propagation direction, the principal $n_i$ and Dirac $\kappa_i$ quantum numbers as well as projection $\mu_i$ of the total angular momentum $j_i = \left| \kappa_i \right| - 1/2$ are used to specify a single-electron bound state. The coupling with the electromagnetic field in this case is taken into account in the second-order perturbation theory, which leads to the matrix element: 
\begin{widetext}
\begin{eqnarray}
    \label{eq:amplitude_general}
    M_{\mu_i m_s}({\bm p}_f) &=& \alpha \sum\limits_{n_\nu \kappa_\nu \mu_\nu} 
    \frac{\mem{{\bm p} m_s}{\hat{R}^\dag({\bm k}_f {\bm \epsilon}_f)}{n_\nu \kappa_\nu \mu_\nu} \mem{n_\nu \kappa_\nu \mu_\nu}{\hat{R}({\bm k}_i {\bm \epsilon}_i)}{n_i \kappa_i \mu_i}}{E_i - E_\nu + \omega_i} \nonumber \\[0.2cm]
    &+& \alpha \sum\limits_{n_\nu \kappa_\nu \mu_\nu} \frac{\mem{{\bm p} m_s}{\hat{R}({\bm k}_i {\bm \epsilon}_i)}{n_\nu \kappa_\nu \mu_\nu} \mem{n_\nu \kappa_\nu \mu_\nu}{\hat{R}^\dag({\bm k}_f {\bm \epsilon}_f)}{n_i \kappa_i \mu_i}}{E_i - E_\nu - \omega_f} \, .
\end{eqnarray}
\end{widetext}
Here, $\alpha$ is the fine structure constant and the summation runs over complete atomic spectrum $\ketm{n_\nu \kappa_\nu \mu_\nu}$, including both bound and (negative and positive) Dirac continuum states. Moreover, photon absorption and emission are described here by the operators $\hat{R}({\bm k}_i {\bm \epsilon}_i)$ and $\hat{R}^\dag({\bm k}_f {\bm \epsilon}_f)$, where
\begin{equation}
   \label{eq:R_operator}
   \hat{R}({\bm k} {\bm \epsilon}) =   {\bm \alpha} \, {\bm \epsilon} \, {\rm e}^{i{\bm k}{\bm r}} \, ,
\end{equation}
and ${\bm \alpha} = \left\{\alpha_x, \alpha_y, \alpha_z \right\}$ is the vector of Dirac matrices. Two terms in the amplitude $M_{\mu_b m_s}({\bm p})$ correspond to respective Feynman diagrams, displayed in Fig.~\ref{Fig2_diagmrams}, and usually referred to as ``absorption first'' and ``emission first'' contributions.

For the evaluation of the amplitude (\ref{eq:amplitude_general}) it is practical to decompose the continuum (emitted) electron state into partial waves:
\begin{eqnarray}
    \label{eq:electron_decomposition}
    \ketm{{\bm p} m_s} &=& \sum\limits_{\kappa_f \mu_f} i^{l_f} \, {\rm e}^{-i \Delta_{\kappa_f}} \, 
    \sqrt{2l_f + 1} \, \sprm{l_f 0 \, 1/2 m_s}{j_f m_s} \nonumber \\
    &\times& D^{j_f}_{\mu_f m_s}(\hat{\bm p}) \, \ketm{E_f \kappa_f \mu_f} \, ,
\end{eqnarray}
which are described by the total energy $E_f$, Dirac quantum number $\kappa_f$ and projection of the total angular momentum $j_f = \left| \kappa_f \right| - 1/2$ on the overall quantization axis, chosen along the propagation direction of incident photons.   By inserting decomposition of $\ketm{{\bm p} m_s}$ into Eq.~(\ref{eq:amplitude_general}) we trivially obtain: 
\begin{eqnarray}
    \label{eq:amplitude_decomposition}
    M_{\mu_i m_s}({\bm p}) &=& \sum\limits_{\kappa_f \mu_f} i^{-l_f} \, {\rm e}^{i \Delta_{\kappa_f}} \, 
    \sqrt{2l_f + 1} \, \sprm{l_f 0 \, 1/2 m_s}{j_f m_s} \nonumber \\
    &\times& D^{j_f *}_{\mu_f m_s}(\hat{\bm p}) \, M_{\mu_i \mu_f}(E_f \kappa_f) \, ,
\end{eqnarray}
where the partial scattering amplitude:
\begin{widetext}
\begin{eqnarray}
    \label{eq:amplitude_partial}
     M_{\mu_i \mu_f}(E_f \kappa_f) &=& \alpha \sum\limits_{n_\nu \kappa_\nu \mu_\nu} 
    \frac{\mem{E_f \kappa_f \mu_f}{\hat{R}^\dag({\bm k}_f {\bm \epsilon}_f)}{n_\nu \kappa_\nu \mu_\nu} \mem{n_\nu \kappa_\nu \mu_\nu}{\hat{R}({\bm k}_i {\bm \epsilon}_i)}{n_i \kappa_i \mu_i}}{E_i - E_\nu + \omega_i} \nonumber \\[0.2cm]
    &+& \alpha \sum\limits_{n_\nu \kappa_\nu \mu_\nu} \frac{\mem{E_f \kappa_f \mu_f}{\hat{R}({\bm k}_i {\bm \epsilon}_i)}{n_\nu \kappa_\nu \mu_\nu} \mem{n_\nu \kappa_\nu \mu_\nu}{\hat{R}^\dag({\bm k}_f {\bm \epsilon}_f)}{n_i \kappa_i \mu_i}}{E_i - E_\nu - \omega_f} \, ,
\end{eqnarray}
\end{widetext}
describes the transition $\ketm{n_i \kappa_i \mu_i} + \gamma_i \to \ketm{E_f \kappa_f \mu_f} + \gamma_f$ between bound and continuum state with well defined total angular momenta and parities. We note that $M_{\mu_i \mu_f}(E_f \kappa_f)$ is very similar to the well-elaborated amplitudes for the Rayleigh and Raman photon scattering. The only difference is that in Compton scattering the final-state electron is no longer bound to the nucleus but belongs to the continuum spectrum. This difference, however, concerns only the radial parts of a scattering amplitude, whereas its angular parts are the same for Rayleigh, Raman and Compton processes. For that reason, we will not further discuss here the evaluation of the partial amplitude (\ref{eq:amplitude_partial}) and refer to Refs.~\cite{JaF14,SuY13,SeS22} for further details. In Sec.~\ref{sec:theory_green} we just briefly review the computations of radial integrals that enter $M_{\mu_i \mu_f}(E_f \kappa_f)$.

Having discussed the second-order matrix element (\ref{eq:amplitude_general}), we are ready to derive the double-differential Compton cross section and the linear polarization of the scattered photons. As mentioned above, a typical experimental scenario where the ejected electron remains unobservable will be considered. For this case, the double-differential cross section
\begin{eqnarray}
   \label{eq:cross_section_S-matrix_1}
   \frac{{\rm d^2}\sigma^{\rm (SM)}}{{\rm d}\Omega_f {\rm d}\omega_f}\left({\bm \epsilon_f}, {\bm \epsilon_i} \right) &=& \frac{1}{2j_i + 1} \nonumber \\[0.2cm]
   && \hspace*{-1.5cm} \times \sum\limits_{\mu_i m_s} \int \left| M_{\mu_i m_s}({\bm \epsilon_f}, {\bm \epsilon_i} ; \, {\bm p}_f) \right|^2 \, {\rm d}\Omega_{{\bm p}_f} \, ,
\end{eqnarray}
can be obtained by integration over the electron emission angles and summation over its polarization. Moreover, we also average over the projection $\mu_i$, assuming that the initial atomic state $\ketm{n_i \kappa_i}$ is unpolarized. By inserting the decomposition (\ref{eq:electron_decomposition}) into Eq.~(\ref{eq:cross_section_S-matrix_1}) and performing standard angular-momentum algebra, one obtains:  
\begin{eqnarray}
   \label{eq:cross_section_S-matrix_2}
   \frac{{\rm d^2}\sigma^{\rm (SM)}}{{\rm d}\Omega_f {\rm d}\omega_f}\left({\bm \epsilon_f}, {\bm \epsilon_i} \right) &=& 2\pi  \sum\limits_{\mu_i \kappa_f \mu_f} \left| M_{\mu_i \mu_f}({\bm \epsilon_f}, {\bm \epsilon_i}; \kappa_f) \right|^2 .
\end{eqnarray}
According to this expression, the double-differential Compton cross section can be written as the sum of (squared) partial scattering amplitudes (\ref{eq:amplitude_partial}). Finally, the cross section (\ref{eq:cross_section_S-matrix_2}), together with Eqs.~(\ref{eq:cross_sections_general}) and (\ref{eq:Stokes_P1_final}), can be employed to obtain the linear polarization of scattered photons.

\section{Green's function approach} 
\label{sec:theory_green}

As follows from the discussion above, the S-matrix calculations of both the double-differential Compton cross section and the polarization of the scattered photons can be traced back to the partial matrix element (\ref{eq:amplitude_partial}). The evaluation of this matrix element is a very challenging task because of the summation and the integration over the \textit{complete} spectrum of virtual intermediate atomic states $\ketm{n_\nu \kappa_\nu \mu_\nu}$. In the present study, this intermediate-state summation and integration is carried out with the help of the relativistic Green's function, defined as:
\begin{equation} \label{eq:G_fct_spectral}
    G({\bm r_2}, {\bm r_1}, E) = \SumInt_{n_\nu \kappa_\nu \mu_\nu} 
    \frac{\phi^{\mu_{\nu}}_{n_\nu \kappa_\nu}({\bm r_2}) \, \phi^{\mu_\nu \dag}_{n_\nu \kappa_\nu}({\bm r_1})}{E_{n_{\nu} \kappa_{\nu}} - E} \, ,
\end{equation}
where the relativistic atomic wavefunctions
\begin{equation}
    \label{eq:wave_function}
    \phi^{\mu_\nu}_{n_\nu \kappa_\nu}({\bm r}) = \frac{1}{r} \left( \begin{array}{c}
    P_{n_\nu \kappa_\nu}(r) \, \chi^{\mu_\nu}_{\kappa_\nu}({\bm \hat{r}})\\[0.2cm]
    i Q_{n_\nu \kappa_\nu}(r) \, \chi^{\mu_\nu}_{-\kappa_{\nu}}({\bm \hat{r}})
    \end{array}\right) \, ,
\end{equation}
can represent bound, positive-continuum, and negative-continuum states. In the continuum case, the quantum number $n_\nu$ should be understood as the total energy of the electron, over which integration is performed. In Eq.~(\ref{eq:wave_function}), moreover, $P_{n_\nu \kappa_\nu}(r)$ and $Q_{n_\nu \kappa_\nu}(r)$ are the radial solutions of the Dirac equation in a central potential, while $\chi^{\mu_\nu}_{\kappa_\nu}({\bm \hat{r}})$ denotes the normalized spin-angular function \cite{Eim95, Eic05}.

By employing the Green's function one can re-write the partial matrix element (\ref{eq:amplitude_partial}) for the Compton scattering as:
\begin{widetext}
\begin{eqnarray}
    \label{eq:amplitude_partial_2}
     M_{\mu_i \mu_f}({\bm \epsilon_f}, {\bm \epsilon_i}; \kappa_f) &=& \alpha \int \phi^{\mu_f \dag}_{E_f \kappa_f}({\bm r}_2) \, {\bm \alpha} \, {\bm \epsilon^*_f} \, {\rm e}^{-i{\bm k}_f{\bm r}_2} \, 
     G({\bm r_2}, {\bm r_1}, E_i + \omega_i) \,  {\bm \alpha} \, {\bm \epsilon_i} \, {\rm e}^{i{\bm k}_i{\bm r}_1} \, \phi^{\mu_i}_{n_i \kappa_i}({\bm r}_1) \, {\rm d}{\bm r}_1 \, {\rm d}{\bm r}_2 \nonumber \\[0.2cm]
     &+& \alpha \int \phi^{\mu_f \dag}_{E_f \kappa_f}({\bm r}_2) \, {\bm \alpha} \, {\bm \epsilon_i} \, {\rm e}^{i{\bm k}_i{\bm r}_2} \, 
     G({\bm r_2}, {\bm r_1},  E_i - \omega_f) \,  {\bm \alpha} \, {\bm \epsilon}^*_f \, {\rm e}^{-i{\bm k}_f{\bm r}_1} \, \phi^{\mu_i}_{n_i \kappa_i}({\bm r}_1) \, {\rm d}{\bm r}_1 \, {\rm d}{\bm r}_2 \, ,
\end{eqnarray}
\end{widetext}
where the intermediate-state summation is ``contained'' now in $G({\bm r_2}, {\bm r_1}, E_i + \omega_i)$ and $G({\bm r_2}, {\bm r_1},  E_i - \omega_f)$. The further evaluation of $M_{\mu_i \mu_f}({\bm \epsilon_f}, {\bm \epsilon_i}; \kappa_f)$ requires the use of the multipole expansions of the electron-photon interaction operator (\ref{eq:R_operator}) and of the Green's function. Since both expansions and the subsequent angular momentum algebra have been discussed at length in the literature, see Refs.~\cite{KoF05,SuK05,SoY24}, we will not elaborate on them here and just mention that Eq.~(\ref{eq:amplitude_partial_2}) can be re-written as (the sum of) products of (i) angular matrix elements and (ii) corresponding radial integrals. While the angular terms are the same as in the theory of Rayleigh and Raman scattering, the radial integrals are different due to the presence of the continuum-state wave function. To discuss this difference and related challenges, we display here one of four radial integrals that enter Eq.~(\ref{eq:amplitude_partial_2}):
\begin{eqnarray}
    \label{eq:radial_integral_1}
    I_{LL} = &\int_0^\infty& {\rm d}r_1 \, {\rm d}r_2 \, 
    Q_{E_f \kappa_f}(r_2) \, j_{L_2}(\omega_{f} r_2) \nonumber \\[0.2cm]
    &\times& g^{LL}_{\kappa_\nu}\left(r_2, r_1; \, E \right) \, j_{L_1}(\omega_i r_1) 
    \, Q_{n_i \kappa_i}(r_1) \, .
\end{eqnarray}
In this expressions, $Q_{n_i \kappa_i}(r_1)$ and $Q_{E_f \kappa_f}(r_2)$ are the ``small'' radial components of the bound- and continuum-state electron wave functions, respectively. The ``large-large'' component of the radial Green's function $g^{LL}_{\kappa_\nu}\left(r_2, r_1; \, E \right)$ is characterized by the Dirac quantum number $\kappa_\nu$ and energy $E$. Moreover, the spherical Bessel functions $j_{L_1}(\omega_i r_1)$ and $j_{L_2}(\omega_{f} r_2)$ arise from the multipole expansion of the electron-photon interaction operator (\ref{eq:R_operator}).

The calculation of the integral (\ref{eq:radial_integral_1}) can be greatly simplified by the fact that the radial component of the Green's function can be written in the form:
\begin{eqnarray}
    \label{eq:radial_Green_function}
    g^{LL}_{\kappa_\nu}(r_2,r_1,E) &=& \frac{1}{w_{\kappa_\nu}} [\Theta(r_2-r_1)F^L_{\kappa_\nu, \infty}(r_2, E) F_{\kappa_\nu, 0}^L(r_1, E) \nonumber\\
    && \hspace{-1cm} + \Theta(r_1-r_2)F^L_{\kappa_\nu,0}(r_2, E) F_{\kappa_\nu, \infty}^L(r_1, E)] \, ,
\end{eqnarray}
where $\Theta$ denotes the Heaviside step function, $F_0^L(r, E)$ and $F^L_\infty(r, E)$ are solutions of the Dirac equation, that are regular at the origin and at infinity, respectively, and $w$ is the corresponding Wronskian. We note that Eq.~(\ref{eq:radial_Green_function}) is valid for an arbitrary central potential. For the particular case of a pure Coulomb interaction, moreover, the exact analytical expressions for $F^{L,S}_{\kappa_\nu, 0}(r)$ and $F^{L,S}_{\kappa_\nu, \infty}(r)$ in terms of Wittaker functions are known and discussed in detail in the literature \cite{SoY24,Hyl84,MoP98}.

By inserting the Green's function (\ref{eq:radial_Green_function}) into Eq.~(\ref{eq:radial_integral_1}) one can write the radial integral $I_{LL}$ in the form:
\begin{eqnarray}
\label{eq:radial_integral_2}
I_{LL} &=& 
\frac{1}{w_{\kappa_\nu}}\int_0^\infty {\rm d}r_1\, Q_{n_i \kappa_i}(r_1)\, F^{L}_{\kappa_\nu,0}(r_1,E)\, j_{L_1}(\omega_i r_1) \nonumber \\[0.2cm]
&& \hspace{-1cm }\times \int_{r_1}^{\infty} dr_2\, Q_{E_f \kappa_f}(r_2)\, F^{L}_{\kappa_\nu,\infty}(r_2,E)\, j_{L_2}(\omega_f r_2)
\nonumber \\[0.3cm]
&&  +
\frac{1}{w_{\kappa_\nu}}\int_0^\infty dr_1\, Q_{n_i \kappa_i}(r_1)\, F^{L}_{\kappa_\nu,\infty}(r_1,E)\, j_{L_1}(\omega_i r_1)
\nonumber \\[0.2cm]
&& \hspace{-1cm }\times
\int_0^{r_1} dr_2\, Q_{E_f \kappa_f}(r_2)\, F^{L}_{\kappa_\nu,0}(r_2,E)\, j_{L_2}(\omega_f r_2)
 \, .
\end{eqnarray}
We recall that for the case of the Compton scattering, considered here, $Q_{E_f \kappa_f}(r_2)$ is the continuum solution of the Dirac equation. Taking into account the asymptotic behavior of this function, as well as that of the radial Green’s function regular at the origin, the integral (\ref{eq:radial_integral_2})  becomes rapidly oscillatory and converges only polynomially. To perform the integration for such a complicated case, we employ two different approaches. The first one follows the method developed in Refs.~\cite{Soy22,SoY23} in which the integration domain is split into regions close to and far from the origin. While one has to perform a \textit{numerical} integration near the origin, the use of asymptotic expansions of Whittaker and spherical Bessel functions for $r_1 \to \infty$ and $r_2 \to \infty$ allows an \textit{analytical} evaluation of the integral at large distances:
\begin{eqnarray}
    \int_0^\infty \; {\rm d}r_1 \, {\mathcal P}(r_1) \int_0^{r_1} \; {\rm d}r_2 \, {\mathcal Q}(r_2) && \nonumber \\[0.2cm]
    && \hspace*{-5cm} = 
    \underbrace{\int_0^{R_0} \; {\rm d}r_1 \, {\mathcal P}(r_1) \int_0^{r_1} \; {\rm d}r_2 \, {\mathcal Q}(r_2)}_\text{Numerical integration} \nonumber\\[0.2cm]
    && \hspace*{-5cm} + \underbrace{\int_{R_0}^{\infty} \; {\rm d}r_1 \, {\mathcal P}(r_1)}_\text{Analytical integration}  \underbrace{\int_0^{R'_0} \; {\rm d}r_2 \, {\mathcal Q}(r_2)}_\text{Numerical integration} \nonumber\\[0.2cm]
    && \hspace*{-5cm} + 
    \underbrace{\int_{R_0}^{\infty} \; {\rm d}r_1 \, {\mathcal P}(r_1) \int_{R'_0}^{r_1} \; {\rm d}r_2 \, {\mathcal Q}(r_2)}_\text{Analytical integration} \, .
\end{eqnarray}
Here, for the sake of brevity we used the short-hand notations ${\mathcal P}(r_1) = \frac{1}{w_{\kappa_\nu}}Q_{n_i \kappa_i}(r_1) \, F^{L}_{\kappa_\nu,\infty}(r_1,E) j_{L_1}(\omega_i r_1)$ and ${\mathcal Q}(r_2) = Q_{E_f \kappa_f}(r_2) \, F^{L}_{\kappa_\nu,0}(r_2,E) \, j_{L_2}(\omega_f r_2)$. Moreover, the parameters $R_0$ and $R'_0$ are chosen such that the arguments of the Whittaker functions appearing in the integrand are sufficiently large, ensuring rapid convergence of their asymptotic expansions. The correctness and convergence of the analytical solutions can be verified by varying $R_0$ and $R'_0$.

As an alternative approach to the evaluation of the integral (\ref{eq:radial_integral_2}), we employ a method based on a Wick rotation of the integration contour. In this method, the asymptotic oscillatory behavior of the integrand, $\sim e^{i(p_f+\omega_f-p_\nu)r}$, is converted into exponential damping by rotating the contour into the negative complex plane, $r \rightarrow -ir$. For $p_\nu > p_f + \omega_f$, the rotated integral converges rapidly and can be evaluated with machine precision. Both approaches yield identical results for the integral, thereby providing a cross-check of the numerical implementation.

\begin{figure*}[t]
    \includegraphics{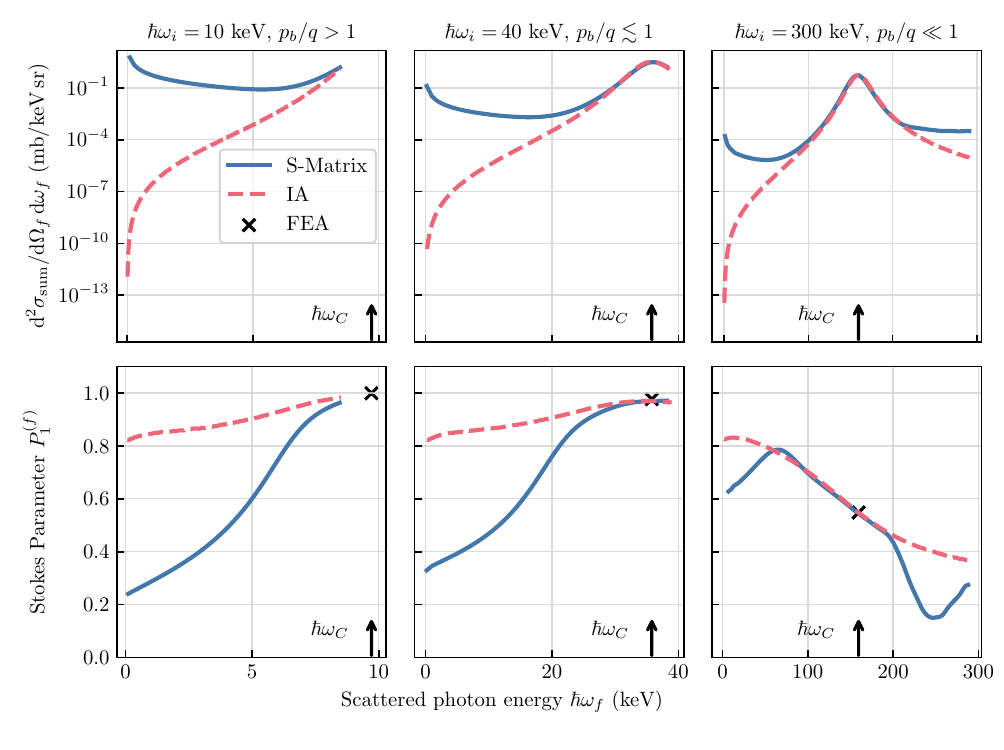}
    \caption{Double-differential Compton cross section ${\rm d}^2\sigma_{\rm sum}$ (upper panels) and linear polarization (lower panels) for photons scattered by hydrogen-like neon, Ne$^{9+}$(1s), at the angle $\theta_f = 120^\circ$. The calculations were performed within the framework of the impulse approximation (IA; red dashed line) and S-matrix theory (blue solid line) for incident photon energies of $\hbar\omega_i = 10$ keV (left column), 40 keV (middle column), and 300 keV (right column). The prediction of the free-electron approximation, Eq.~(\ref{eq:Stokes_P1_final_FEA}), is also shown (FEA; cross symbol) for the polarization of the scattered photons.}
     \label{Fig3}
\end{figure*}
\begin{figure*}[t]
    \includegraphics{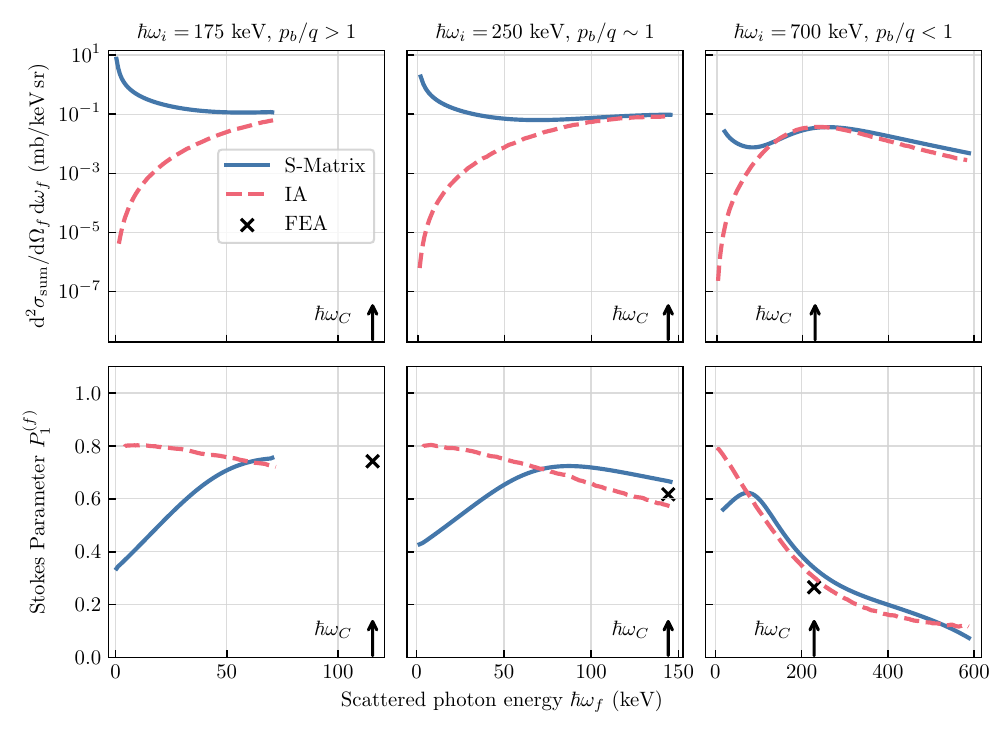}
    \caption{Double-differential Compton cross section ${\rm d}^2\sigma_{\rm sum}$ (upper panels) and linear polarization (lower panels) for photons scattered by hydrogen-like lead, Pb$^{81+}$(1s), at the angle $\theta_f = 120^\circ$. The calculations were performed within the framework of the impulse approximation (IA; red dashed line) and S-matrix theory (solid line) for incident photon energies of $\hbar\omega_i = $ 175 keV (left column), 250 keV (middle column) and 700 keV (right column). For the polarization of the scattered photons, the prediction of the free-electron approximation, Eq.~(\ref{eq:Stokes_P1_final_FEA}), is also shown (FEA; cross symbol).}
     \label{Fig4}
\end{figure*}
\section{Results and discussion} 
\label{sec:results}

After reviewing the theoretical background, we are ready to present the results of our calculations for the double-differential Compton cross section and the linear polarization of the scattered photons. We start with the scattering of completely linearly polarized $X$ rays with energies $\hbar\omega_i = 10$~keV, 40~keV, and 300~keV by a hydrogen-like neon ion Ne$^{9+}$, initially in the ground $1s$ state. In Fig.~\ref{Fig3} we compare the impulse-approximation (red dashed line) and $S$-matrix (blue solid line) predictions for the cross section (upper panels) and the Stokes parameter $P^{(f)}_1$ (lower panels). Whereas $P^{(f)}_1$ is obtained from Eq.~(\ref{eq:Stokes_P1}), the cross section corresponds to the case in which the polarization of the scattered photons is not analyzed:
\begin{eqnarray}
    \frac{{\rm d}^2\sigma_{\rm sum}}{{\rm d}\Omega_f {\rm d}\omega_f}\left(P_1^{(i)}\right) &=& \sum\limits_{\epsilon_f} \frac{{\rm d}^2\sigma}{{\rm d}\Omega_f {\rm d}\omega_f}\left({\bm \epsilon_f},  P_1^{(i)}\right) \, .
\end{eqnarray}
In what follows, we refer to this quantity as the ``polarization-summed'' cross section  and denote it by ${\rm d}^2\sigma_{\rm sum}$ for shortness. Both, ${\rm d}^2\sigma_{\rm sum}$ and $P^{(f)}_1$ are displayed here as functions of final photon energy $\hbar \omega_f$ and for the scattering angle $\theta_f = 120^\circ$. The IA and S-matrix polarization calculations are compared, moreover, with the free-electron result (\ref{eq:Stokes_P1_final_FEA}), which is applicable for the Compton energy $\hbar \omega_C$. 

The three incident photon energies $\omega_i$ selected for the calculations in Fig.~\ref{Fig3} allow us to probe different kinematic regimes of Compton scattering. These regimes can be discussed from two complementary viewpoints, namely in terms of energy and momentum. From the energy viewpoint, the kinematic regimes can be characterized by the ratio $\omega_i / \omega_C$, which quantifies the shift of the Compton peak and thus reflects the roil of recoil. Using Eq.~(\ref{eq:Compton_energy}), one finds that, for a fixed scattering angle, this ratio grows \textit{linearly} with the incident energy:
\begin{equation}
   \frac{\omega_i}{\omega_C} = 1 + \omega_i \left(1 - \cos\theta_f\right) \, ,
\end{equation}
As $\omega_i$ increases, the calculations in Fig.~\ref{Fig3} span different recoil regimes, ranging from $\omega_i \ll m_e$, where recoil is weak, to $\omega_i \sim m_e$, where recoil effects become substantial. This evolution is reflected in the strong leftward shift of the Compton peak (black arrow), from $\omega_C \approx \omega_i$ at 10~keV to $\omega_C \approx 0.53\,\omega_i$ at 300~keV. 

From the momentum viewpoint, the kinematic regimes explored in Fig.~\ref{Fig3} are governed by the ratio of the characteristic bound-electron momentum $p_b$ to the photon momentum transfer $q$. For an incident photon energy of 300~keV, the ratio is much smaller than unity, $p_b/q \ll 1$, indicating that the binding potential becomes negligible during the scattering, and the electron effectively behaves as a quasi-free particle. The use of the impulse approximation is therefore well justified in this regime, as can be seen from the right column of Fig.~\ref{Fig3}. Indeed, both the IA and the more sophisticated S-matrix approach predict double-differential cross sections that are in very good agreement over a fairly wide energy range around the Compton peak, $\hbar\omega_C \approx 160$~keV. Moreover, the IA and S-matrix calculations for the linear polarization of the scattered photons $P_1^{(f)}$ also match well with each other for 80~$\lesssim \hbar \omega_f \lesssim$~200 keV. In addition, at the position of the Compton peak, the predictions of both approaches reproduce the free-electron result, $P_1^{(f)} \approx 0.55$, at the 1~\% level.  

\begin{figure*}[t]
    \includegraphics{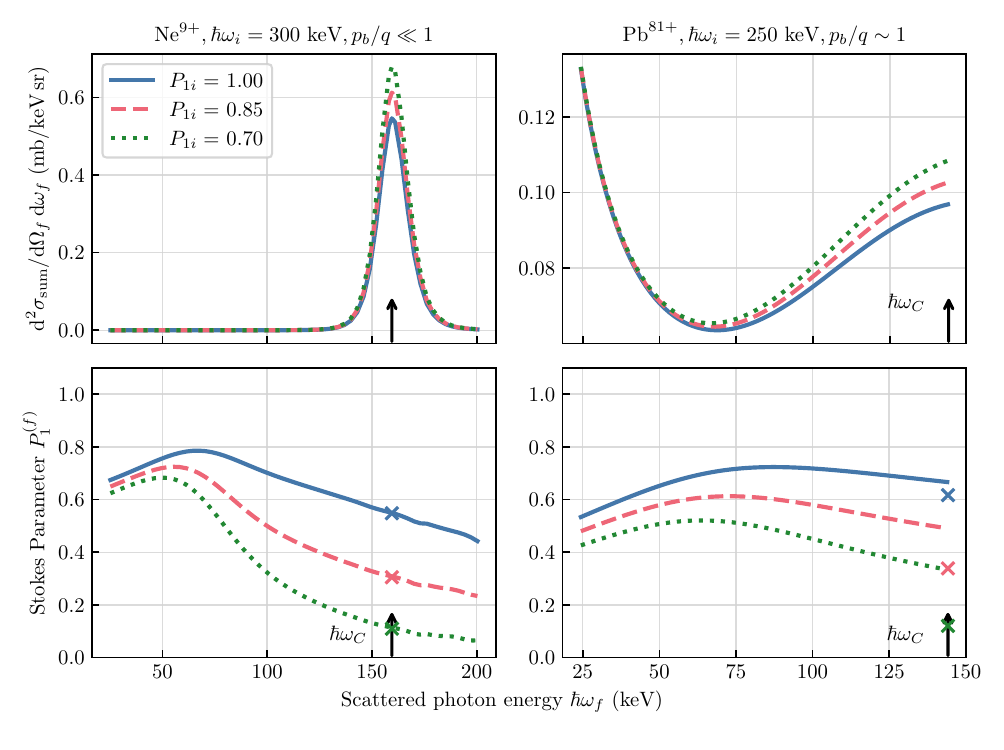}
    \caption{Double-differential Compton cross section ${\rm d}^2\sigma_{\rm sum}$ (upper panels) and linear polarization (lower panels) for 300~keV photons scattered by Ne$^{9+}$ (left column) and 250~keV photons scattered by Pb$^{81+}$ (right column), at the angle $\theta_f = 120^\circ$. The S-matrix calculations were performed for three degrees of linear polarization of the incident radiation: $P_1^{(i)} = 1$ (blue solid line), $P_1^{(i)} = 0.85$ (red dashed line), and $P_1^{(i)} = 0.7$ (green dotted line). The predictions of the free-electron approximation, Eq.~(\ref{eq:Stokes_P1_final_FEA}), are also shown (cross symbol).}
     \label{Fig5}
\end{figure*}

While the impulse approximation provides a good description of both the cross section ${\rm d}^2\sigma_{\rm sum}$ and the polarization $P^{(f)}_1$ for $\hbar\omega_i = 300$~keV, the situation changes noticeably as the incident photon energy decreases. In this case, the ratio $p_b/q$ increases and may even exceed unity, indicating that the electron binding effects play an important role in the scattering process. The middle and left columns of Fig.~\ref{Fig3} clearly show that as $p_b/q$ becomes larger, the agreement between the IA and S-matrix results rapidly becomes restricted to a narrow interval around the Compton peak. In particular, for $\hbar \omega_f < \hbar \omega_C$ the impulse approximation strongly underestimates the polarization-summed cross section ${\rm d}^2\sigma_{\rm sum}$ and fails to predict the onset of its infrared divergence at $\hbar\omega_f \to 0$, a behavior discussed in detail by Bergstrom and co-workers in Ref.~\cite{BeS93}. Moreover, the S-matrix calculations predict a much stronger depolarization of the outgoing radiation compared with the IA results. This depolarization is most pronounced at low final photon energies $\hbar\omega_f$, where Compton scattering becomes increasingly sensitive to electron-binding effects neglected in the impulse approximation.

Having discussed Compton scattering by the $K$-shell of neon, we now proceed to the case of lead, Pb$^{81+}$, as an example of a heavy atomic target. Interest in such high-$Z$ targets irradiated by polarized X-rays has been stimulated by recent experiments at the PETRA III synchrotron at DESY \cite{BlF16}. In Fig.~\ref{Fig4}, we display the double-differential cross section ${\rm d}^2\sigma_{\rm sum}$ and the linear polarization $P^{(f)}_1$ of photons scattered by Pb$^{81+}$(1s) at the angle $\theta_f = 120^\circ$. The calculations were carried out within the impulse approximation and the S-matrix theory for linearly polarized incident photons with energies $\hbar\omega_i$ = 175~keV (left column), 250~keV (middle column) and 700~keV (right column). As in the case of neon discussed above, the three incident energies may also be viewed from the energy perspective as probing different recoil regimes. In the following, however, we focus primarily on their momentum-space interpretation in terms of the ratio $p_b/q$, since this parameter is more directly relevant for comparing the IA and S-matrix results. For the higher photon energy $\hbar \omega_i$~=~700~keV, for example, this ratio is less than unity, suggesting that the impulse approximation should provide a reasonable description of the scattering process. Indeed, as shown in the right column of Fig.~\ref{Fig4}, the IA and S-matrix calculations exhibit the same overall energy dependence and remain close to each other over the interval $100 \lesssim \hbar\omega_f \lesssim 500$~keV around the Compton peak $\hbar \omega_C = 230$~keV, both for the differential cross section and for the polarization $P^{(f)}_1$. However, a markedly different behavior emerges for the lower incident photon energy $\hbar\omega_i = 175$~keV, corresponding to the regime $p_b/q > 1$. Here, no Compton peak appears in the double-differential cross section, since the large binding energy of the $K$-shell electron in Pb$^{81+}$ limits the scattered-photon energy $\hbar\omega_f$ to values below $\hbar\omega_C$. Moreover, as expected, the impulse-approximation and S-matrix calculations remarkably disagree in this kinematic regime. In particular, as in the case of the Ne target discussed above, the IA yields differential cross sections that are significantly smaller than the corresponding S-matrix predictions. In contrast, for the linear polarization of the scattered photons, the impulse-approximation calculations overestimate the S-matrix results, except for photon energies $\hbar\omega_f \approx 60$--$70$~keV.

\begin{figure*}[t]
    \includegraphics{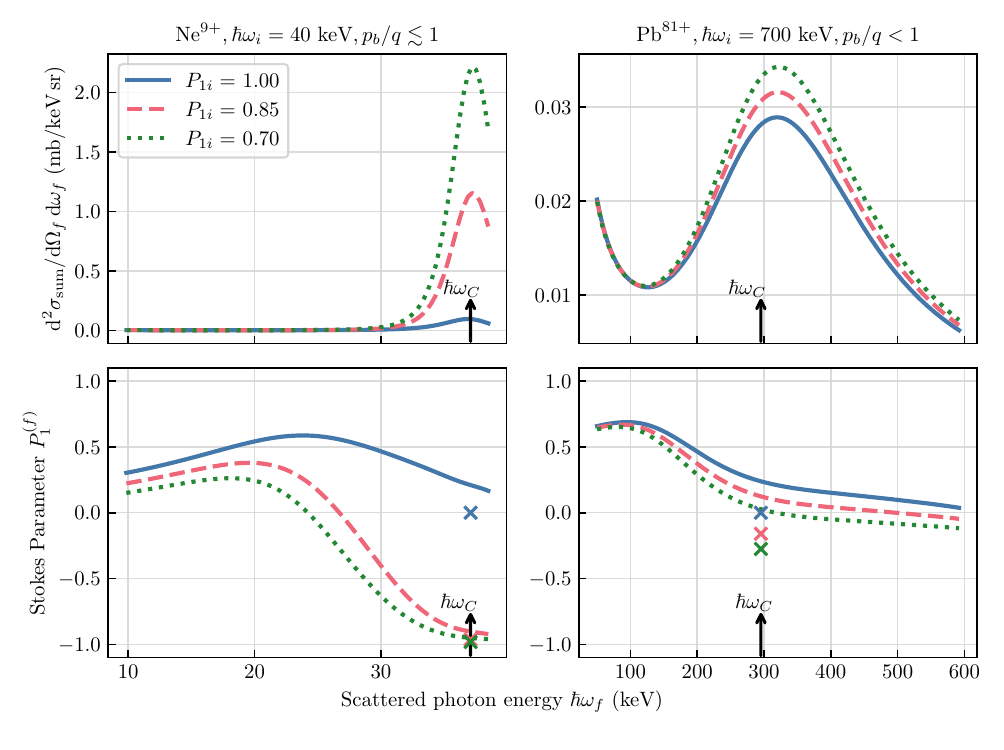}
    \caption{Double-differential Compton cross section ${\rm d}^2\sigma_{\rm sum}$ (upper panels) and linear polarization (lower panels) for 40~keV photons scattered by Ne$^{9+}$ (left column) and 700~keV photons scattered by Pb$^{81+}$ (right column), at the angle $\theta_f = 90^\circ$. The S-matrix calculations were performed for three degrees of linear polarization of the incident radiation: $P_1^{(i)} = 1$ (blue solid line), $P_1^{(i)} = 0.85$ (red dashed line), and $P_1^{(i)} = 0.7$ (green dotted line). The predictions of the free-electron approximation, Eq.~(\ref{eq:Stokes_P1_final_FEA}), are also shown (cross symbol).}
     \label{Fig6}
\end{figure*}

So far, we have discussed Compton scattering of $X$-rays that are completely linearly polarized in the reaction plane, $P_1^{(i)} = 1$. As mentioned above, this is not the case for recent synchrotron experiments, in which the degree of polarization of incident radiation is slightly below unity \cite{BlF16}. In what follows, therefore, we employ the S-matrix theory to investigate how the differential cross section ${\rm d}^2\sigma_{\rm sum}$ and the polarization of the scattered photons $P_1^{(f)}$ depend on $P_1^{(i)}$. For instance, Fig.~\ref{Fig5} displays results for the scattering of 300~keV photons by Ne$^{9+}$ (left column) and 250~keV photons by Pb$^{81+}$ (right column). The S-matrix calculations were performed for the scattering angle $\theta_f = 120^\circ$ and for several degrees of linear polarization of the incident radiation: $P_1^{(i)} = 1$ (blue solid line), $P_1^{(i)} = 0.85$ (red dashed line), and $P_1^{(i)} = 0.7$ (green dotted line). As seen from the figure, the reduction of $P_1^{(i)}$ affects both the Compton cross section and the polarization of the scattered photons. In particular, one observes that ${\rm d}^2\sigma_{\rm sum}$ becomes larger, whereas $P_1^{(f)}$ becomes smaller, as the degree of linear polarization $P_1^{(i)}$ of the incident radiation decreases. To gain a \textit{qualitative} understanding of this behaviour, we refer to the classical Thomson limit of Compton scattering (\ref{eq:Thomson}), in which the polarization-resolved cross section depends directly on the relative orientation of the initial- and final-state polarization vectors. For the geometry considered in the present study (see Fig.~\ref{Fig1_geometry}), this leads to the following expression for the cross section describing the scattering of partially polarized incident radiation, after summing over the polarization of the outgoing photons:
\begin{equation}
\frac{{\rm d}^2\sigma^{\rm (Th)}_{\rm sum}}{{\rm d}\Omega_f {\rm d} \omega_f}\left(P_1^{(i)} \right) \propto \frac{1+\cos^2\theta_f}{2} - \frac{P_1^{(i)}}{2}\sin^2\theta_f \, .
\end{equation}
The above relation clearly indicates that, as $P_1^{(i)}$ decreases from unity while remaining positive, the cross section increases. This behavior has a simple classical interpretation. For fully linearly polarized incident radiation with the polarization vector lying in the scattering plane, i.e. for $P_1^{(i)} = 1$, scattering in this plane is suppressed, since the electron oscillates along the polarization direction and dipole radiation vanishes along the oscillation axis. As the incident radiation becomes less polarized, $P_1^{(i)} < 1$, a perpendicular polarization component contributes increasingly to the scattering, thereby reducing this suppression and enhancing the cross section \cite{BeS93}.

In contrast to the cross section, the degree of linear polarization $P_1^{(f)}$ of the scattered photons is governed by the \textit{difference} between the polarization-resolved cross sections ${\rm d}^2\sigma\left({\bm \epsilon}_{f,\parallel}, P_1^{(i)}\right)$ and ${\rm d}^2\sigma\left({\bm \epsilon}_{f,\perp}, P_1^{(i)}\right)$, normalized to their sum, see Eq.~(\ref{eq:Stokes_P1_final}). In the Thomson limit the difference is given by:
\begin{eqnarray}
\frac{{\rm d}^2\sigma^{\rm (Th)}}{{\rm d}\Omega_f {\rm d} \omega_f}\left({\bm \epsilon}_{f,\parallel},P_1^{(i)}\right)
-
\frac{{\rm d}^2\sigma^{\rm (Th)}}{{\rm d}\Omega_f {\rm d} \omega_f}\left({\bm \epsilon}_{f,\perp},P_1^{(i)}\right)
&& \nonumber \\[0.2cm]
&& \hspace*{-5cm} \propto \frac{\cos^2\theta_f - 1}{2}
+
\frac{1+\cos^2\theta_f}{2}\,P_1^{(i)} \, .
\end{eqnarray}
This expression implies that an incident beam, fully linearly polarized along the $x$--axis ($P_1^{(i)} = 1$), gives rise to the largest difference between the cross sections ${\rm d}^2\sigma^{\rm (Th)}\!\left({\bm \epsilon}_{f,\parallel}, P_1^{(i)}=1\right) \propto \cos^2\theta_f$ and ${\rm d}^2\sigma^{\rm (Th)}\!\left({\bm \epsilon}_{f,\perp}, P_1^{(i)}=1\right) = 0$. As $P_1^{(i)}$ decreases, the difference becomes smaller, which in turn leads to a reduction of the polarization $P_1^{(f)}$ of the scattered photons.  In terms of the classical dipole picture, a reduction of the incident linear polarization $P_1^{(i)}$ weakens the preferred oscillation direction of an electron, induced by the incoming field, and thereby diminishes the difference between the two orthogonal polarization components of the scattered radiation. Thus, despite the additional complexity introduced by relativistic and binding effects, the Thomson-limit picture provides a simple qualitative explanation of the very $P_1^{(i)}$-dependence observed in Fig.~\ref{Fig5}.

The depolarization of the scattered photons as the parameter $P_1^{(i)}$ decreases is also predicted by the free-electron approximation. In the kinematic regime where the bound-electron momentum is small compared with the momentum transfer, $p_b/q \ll 1$, and the scattered-photon energy $\hbar\omega_f$ is in the vicinity of the Compton peak, the FEA and the S-matrix results exhibit not only qualitative but also quantitative agreement. This is illustrated in the left column of Fig.~\ref{Fig5}, where the results for $K$-shell scattering on Ne$^{9+}$ are presented. Indeed, both theories predict at the Compton peak $P_1^{(f)} \approx 0.55$, $0.30$, and $0.10$ for $P_1^{(i)} = 1$, $0.85$, and $0.7$, respectively. In contrast, for the scattering of 250~keV X-rays by Pb$^{81+}$ (right column of the figure), the FEA no longer reproduces the S-matrix results quantitatively. Such a behavior is expected since the kinematic parameter $p_b/q$ is of the order of unity in the present case, indicating that electron binding effects play an important role. For Pb$^{81+}$, these effects are further enhanced by the large nuclear charge, which leads to a strongly bound electronic state and hence to a larger deviation from the free-electron picture. Nevertheless, the FEA still captures the overall trend of the results: the depolarization of the scattered photons with decreasing incident polarization $P_1^{(i)}$.

While in the previous figures we examined Compton scattering at an angle $\theta_f = 120^\circ$, Fig.~\ref{Fig6} presents results for $\theta_f = 90^\circ$. This is a special geometry in which scattering in the plane of linear polarization of the incident radiation is strongly suppressed. In particular, in the classical dipole (Thomson) limit the cross section (\ref{eq:Thomson}) vanishes identically when ${\bm \epsilon}_i$ and ${\bm \epsilon}_f$ are orthogonal. Consequently, scattering at $90^\circ$ is highly sensitive to the polarization of the incident radiation: even small variations of $P_1^{(i)}$ may lead to pronounced changes in both the intensity and the polarization of the outgoing photons. In the past, the high polarization sensitivity of elastic Rayleigh scattering at angles near $90^\circ$ has been used to diagnose the polarization of radiation from the PETRA~III synchrotron \cite{BlF16,MiG26}. A similarly pronounced sensitivity to $P_1^{(i)}$ is observed for Compton scattering of 40~keV X-rays by Ne$^{9+}$, as seen in the left column of Fig.~\ref{Fig6}. The polarization of the scattered photons with energies near the Compton peak even changes sign for a relatively small decrease of $P_1^{(i)}$ from 1.0 to 0.85. Such a strong $P_1^{(i)}$-dependence exhibited in the Ne$^{9+}$ case can be explained by the fact that the kinematic condition $\hbar \omega_f \approx \hbar\omega_i$ holds, implying that the scattering approaches the Thomson limit. As discussed above, in this limit and for $\theta_f = 90^\circ$, the scattering channel in which both the incident and scattered photons are polarized within the reaction plane is strongly suppressed. As a result, even a relatively small fraction of incident photons, polarized perpendicular to the reaction plane, can dominate the scattered intensity and thus lead to $P_1^{(f)}<0$.

In contrast to the Ne$^{9+}$ case, the cross section and polarization of the photons scattered by Pb$^{81+}$ are less sensitive to the polarization of the incident 700~keV $\gamma$-rays, see the right column of Fig.~\ref{Fig6}. The weaker polarization dependence can be attributed to the significant recoil, $\hbar\omega_f < \hbar\omega_i$, which indicates that the scattering takes place far from the Thomson regime. Both effects reduce the sensitivity of the scattering process to $P_1^{(i)}$.

\section{Summary and outlook} 
\label{sec:summary}

In summary, we performed a theoretical study of Compton scattering of linearly polarized X- and $\gamma$-rays by bound electrons. Particular emphasis was placed on the double-differential cross section and the polarization of the scattered photons for the case when the emitted  electron is not observed. Both properties were analyzed using the S-matrix theory and the relativistic Green's function approach. Moreover, predictions of the free-electron and impulse approximations were also obtained, providing an additional benchmark for our calculations.

Based on the developed theoretical approaches, detailed calculations have been performed for Compton scattering by a $K$-shell electron in hydrogen-like Ne$^{9+}$ and Pb$^{81+}$ ions. A broad range of incident photon energies and scattering angles was considered to probe various kinematic regimes. These regimes can be discussed from two complementary viewpoints, namely in terms of energy and momentum. For the former, the relevant quantity is the ratio $\omega_i/\omega_C$, which quantifies the shift of the Compton peak and thus reflects the role of recoil. For the latter, one considers the ratio of the bound-electron momentum to the photon momentum transfer, $p_b/q$, which indicates how strongly binding affects the scattering. Since the momentum-based viewpoint is commonly used to assess the validity of the impulse approximation, we mainly rely on it in our analysis. In particular, for $p_b/q \lesssim 1$ and energies near the Compton peak, the S-matrix results generally agree well with the impulse-approximation predictions. For $p_b/q > 1$, where electron binding becomes increasingly important, the present calculations indicate that the impulse approximation usually overestimates the polarization of the scattered light and underestimates the Compton cross section. Both deviations become more pronounced as one moves away from the Compton peak.

Of particular interest in the present work is the dependence of the Compton cross section and, especially, the polarization of the scattered photons on the degree of linear polarization of the incident radiation. This investigation was motivated by recent photon-scattering experiments at the PETRA~III facility, where the synchrotron radiation was slightly depolarized. Our calculations show that, for the kinematic conditions explored in the present work, a reduction of the incident linear polarization leads to a marked depolarization of the scattered photons, accompanied by an increase in the Compton cross section. A particularly pronounced polarization effect occurs for scattering at angles near $90^\circ$ and in the small-recoil regime, i.e. when $\hbar\omega_f \approx \hbar\omega_i$.  In the cases studied here, this effect is especially strong for low-$Z$ targets, for which binding effects are weaker and the scattering more closely approaches the Thomson limit. Moreover, the present calculations demonstrate that a quantitative analysis of polarization in bound-state Compton scattering over a wide range of kinematic regimes requires a rigorous $S$-matrix treatment rather than the simplified free-electron and impulse-approximation approaches.

The present theoretical study has been restricted to Compton scattering by hydrogen-like ions. Hence, both the wave- and the Green's functions of an electron were generated for a pure Coulomb potential. One can note that one-electron atomic systems represent, of course, a special case. Nevertheless, X- and $\gamma$-ray scattering by heavy hydrogen-like ions is of interest for future experiments within the Gamma Factory project at CERN \cite{BuC20} and at the FAIR facility in Darmstadt. Moreover, test calculations, aimed at reproducing the data of Bergstrom \textit{et al.} \cite{BERGSTROM19973}, show that Coulombic wave- and Green's functions provide reasonably accurate predictions also for inner-shell scattering in neutral atoms. Still, electron screening effects in the Compton scattering by heavy many-electron atoms require particular attention. In future work, we aim to extend the S-matrix theory to include mean-field screening potentials and apply it to the detailed analysis of polarization effects in the Compton scattering.

\section*{Acknowledgements} 

This work was supported by the German Research Foundation (Deutsche Forschungsgemeinschaft, DFG) under project SO 2403/2-1.

J.S. and N.M.M. contributed equally to this paper.


\bibliography{./a24.references}

\end{document}